\documentclass[10pt,conference]{IEEEtran}
\IEEEoverridecommandlockouts

\usepackage[numbers,sort&compress]{natbib}
\usepackage{amsmath,amsthm,amssymb,amsfonts}
\usepackage{bbm}
\usepackage{algorithm}
\usepackage{algorithmic}
\usepackage{graphicx}
\usepackage{textcomp}
\usepackage{xcolor}
\usepackage{flushend}
\usepackage{hyperref}
\hypersetup{
  colorlinks=false,
  pdfborder={0 0 0}
}
\usepackage[numbers]{natbib}

\usepackage{subfig}
\usepackage{algorithm}
\usepackage[font=small]{caption}

\def\BibTeX{{\rm B\kern-.05em{\sc i\kern-.025em b}\kern-.08em
    T\kern-.1667em\lower.7ex\hbox{E}\kern-.125emX}}
\begin{document}

\title{TopoEdge: Topology-Grounded Agentic Framework for Edge Networking Code Generation and Repair\\
}

\author{
    \IEEEauthorblockN{Haomin Qi$^{1}$, Bohan Liu$^{2}$, Zihan Dai$^{3}$, Yunkai Gao$^{4}$}
    \IEEEauthorblockA{
        $^{1}$University of California San Diego, La Jolla, CA, USA \\
        $^{2}$The Chinese University of Hong Kong, Hong Kong SAR \\
        $^{3}$University of Copenhagen, Copenhagen, Denmark \\
        $^{4}$Duke University, Durham, NC, USA \\
        Email: h5qi@ucsd.edu, 1155173765@link.cuhk.edu.hk, cjh841@alumni.ku.dk, yunkai.gao@duke.edu
    }
}

\maketitle

\begin{abstract}
TopoEdge is a topology-grounded, edge-deployable framework for end-to-end software-defined networking (SDN) configuration generation and repair, motivated by the brittleness of configuration artefacts under topology variation and by strict operational constraints on latency, privacy, and on-site execution. TopoEdge represents each target topology as a router-level graph and embeds it using a contrastively trained graph neural network (GNN), enabling nearest-neighbour retrieval of a verified reference configuration paired with an executable Python driver (a Topotest/pytest test script that orchestrates the emulated network and checks protocol assertions). The target topology, retrieved reference topology, and reference driver are assembled into a topology-grounded retrieval-augmented generation context (TopoRAG), which grounds a distributed, execution-centric generate--verify--repair loop coordinated by a central controller and realised by three role-specialised agents: (i) a Planning agent that produces a topology-consistent configuration plan and a per-device skeleton; (ii) a Generation agent that materialises executable configuration artefacts, including device configurations and the driver; and (iii) a Verification agent that runs the FRRouting Topotest/pytest harness, compresses failures into a compact trace, and emits localised patch directives for iterative repair.
\end{abstract}


\section{INTRODUCTION}
\label{sec:introduction}
Software-defined networking (SDN) enables fine-grained control over routing, addressing, and policy enforcement, but it also expands the operational surface area of configuration. As networks scale, configuration logic must remain consistent with topology, protocol interactions, and device-level constraints, and small mistakes can manifest as silent semantic failures that are difficult to diagnose \cite{mondal2025ambiguity}. Existing automation tools help with syntax and templating, yet they often lack explicit topology awareness and rarely close the loop with execution feedback. At the same time, many operators prefer on-site inference for privacy, security, and cost reasons, which motivates designs that work with small models on resource-constrained edge hardware rather than relying on monolithic centralised inference \cite{wang2024netconfeval}.

TopoEdge is built around the observation that topology is a stable signal for transferring configuration patterns. It introduces TopoRAG, which learns a structure-sensitive embedding space over FRRouting topologies \cite{FRRTopotests} and retrieves a verified nearest neighbour case with a runnable driver for ground generation. Given a target topology, TopoEdge parses the JSON into a router-level graph, embeds it with a contrastively trained three-layer graph convolutional network (GCN), and performs cosine nearest neighbour retrieval over a reference set of verified cases. The retrieved topology and driver, together with the target topology and SDN background knowledge, form the shared context for a generate-verify-repair loop. This loop is executed by three role-specialised AI agents: Planning agent, Generation agent, and Verify agent. Planning produces a topology-consistent protocol plan and device skeleton. Generation materialises executable artefacts, and Verify runs the FRRouting Topotest harness with pytest, trims failures into compact traces, and issues localised patch directives for iterative repair. Two lightweight controllers further improve edge practicality: an adaptive inference budget controller allocates per-case token and iteration limits using simple difficulty signals such as topology size and retrieval similarity, while a constrained decoding layer restricts generation to schema-valid parameters and a curated command lexicon so that the loop spends iterations on semantic corrections rather than syntactic recovery.

We make two contributions. First, we propose TopoRAG, a topology-aware retrieval module that aligns nearest neighbour topology similarity with verified SDN drivers, providing transferable structure and protocol intent for downstream generation. Second, we propose TopoEdge, an edge-deployable agentic framework that uses TopoRAG grounding and an execution-centric generate–verify–repair loop to generate and repair SDN configurations under resource constraints. In our results on 200 held-out query cases, TopoEdge reaches Pass@20 of 0.890, substantially higher than No-TopoRAG at 0.550 and close to the centralised large language model (LLM) baseline at 0.930, while maintaining an edge-deployable design.

\section{RELATED WORK}
\label{sec:relatedwork}
\subsection{Automated SDN Configuration}

Early work on SDN reliability applies model checking, data-plane verification, and synthesis to validate reachability, isolation, and waypointing on abstract control- and data-plane models. These methods can detect subtle issues before deployment, but typically assume that configuration artefacts or policies already exist and are often separated from day-to-day authoring of vendor-specific scripts.

Large language models have been evaluated as configuration generators. NETCONF benchmarks network configuration and shows that exact syntax, topology awareness, and long-range constraints remain challenging without careful prompting and tooling. Complementary work on low-level NETCONF generation demonstrates realistic device outputs but emphasises strict validation and context control \cite{hollosi2024netconfnetconf}. In zero-touch management, LLM-NetCFG builds configuration agents that translate intents into device configurations and verify them, using a local LLM for privacy and control \cite{lira2024llmnetcfg}. These systems demonstrate promise, but they largely rely on centralised inference, treat configuration as text-centric, and do not integrate explicit topology-driven retrieval or an edge-oriented multi-agent loop with adaptive reasoning budgets.

TopoEdge targets routine configuration and repair by grounding generation in topology and closing the loop with execution feedback. It couples topology-aware retrieval over verified reference cases with an agentic generate--verify--repair loop deployed under resource constraints.

\subsection{Topology-aware Representations for Network Management}

Topological deep learning extends graph learning to simplicial and cellular complexes, enabling models to capture higher-order motifs such as paths and cycles \cite{papillon2023architectures}. This motivates structure-aware encoders for networked systems.

In our setting, sequence-based encoders over linearised JSON or code, including multilayer perceptrons (MLPs), autoencoders, and Bidirectional Encoder Representations from Transformers (BERT)-style models, capture local syntax and token statistics but are weak at separating cases that share similar text while differing in router-level connectivity. We also explored topological neural network toolchains that lift graphs into higher-order complexes \cite{telyatnikov2024topobenchmarkx}. However, the FRRouting topologies 
\cite{FRRTopotests} used here are largely rank-one, consisting of routers and switches with links, so higher-order machinery adds limited discriminative value over graphs, and practical network-modeling studies report that current graph neural networks (GNN) families can still exhibit systematic limitations under realistic network structure variation \cite{happ2022gnnlimitations, qi2026graphcue}. In addition, pipelines such as the Simplicial Attention Network in TopoModelX require conversion into opaque pickle-based objects, increasing operational friction for debugging and data inspection. For lightweight retrieval that must integrate cleanly with an edge-deployed loop, these costs are difficult to justify.

TopoEdge therefore adopts a conventional graph convolutional encoder on router-level graphs, preserving the structural cues needed for nearest-neighbour retrieval while keeping computation and data formats simple.

\subsection{Edge AI and Agentic AI Systems}

Efficient LLM surveys catalogue quantisation, pruning, distillation, parameter-efficient fine-tuning, and speculative decoding, highlighting accuracy--memory--latency tradeoffs \cite{wan2023efficientllm}. Edge deployment discussions further argue that constrained hardware motivates careful system design and, when necessary, offloading and partitioning strategies. In networking, surveys and empirical studies show LLM utility for intent translation, troubleshooting, and tool-augmented closed-loop control in testbeds \cite{siino2025testbeds,hollosi2024netconfnetconf}. Agentic frameworks extend this direction by decomposing workflows into specialised roles.

TopoEdge is distinguished by combining three elements that are rarely addressed jointly: topology-aware retrieval grounded in verified SDN instances, distributed edge deployment of quantised small models, and explicit control of inference cost via adaptive budgets and constrained decoding. This combination yields a concrete, resource-aware design for reproducible SDN configuration and verification on edge hardware.

\section{FRAMEWORK DESIGN}
\label{sec:framework}


\subsection{Topology Model Training and TopoRAG}\label{subsec:topo_training}

Each Topotest case provides a topology JSON that lists devices, their interfaces, and the links between them. TopoEdge parses each JSON into an undirected graph \(G=(V,E,X)\), where \(V\) is the set of nodes, \(E \subseteq V \times V\) is the set of undirected edges, and \(X \in \mathbb{R}^{|V| \times F}\) is the node feature matrix. Nodes correspond to routers and switches. An edge \((u,v)\) exists if the JSON contains a link between devices \(u\) and \(v\). Each node feature vector \(x_i \in \mathbb{R}^{F}\) concatenates a one-hot encoding of the device type with simple structural attributes such as node degree and a lightweight configuration counter, yielding a compact feature space with \(F=4\). This representation deliberately emphasises router-level connectivity and coarse structural cues, because these signals are stable across cases and are most useful for retrieving configuration patterns that transfer.

\begin{figure*}[t]
  \centering
  \includegraphics[width=0.9\textwidth]{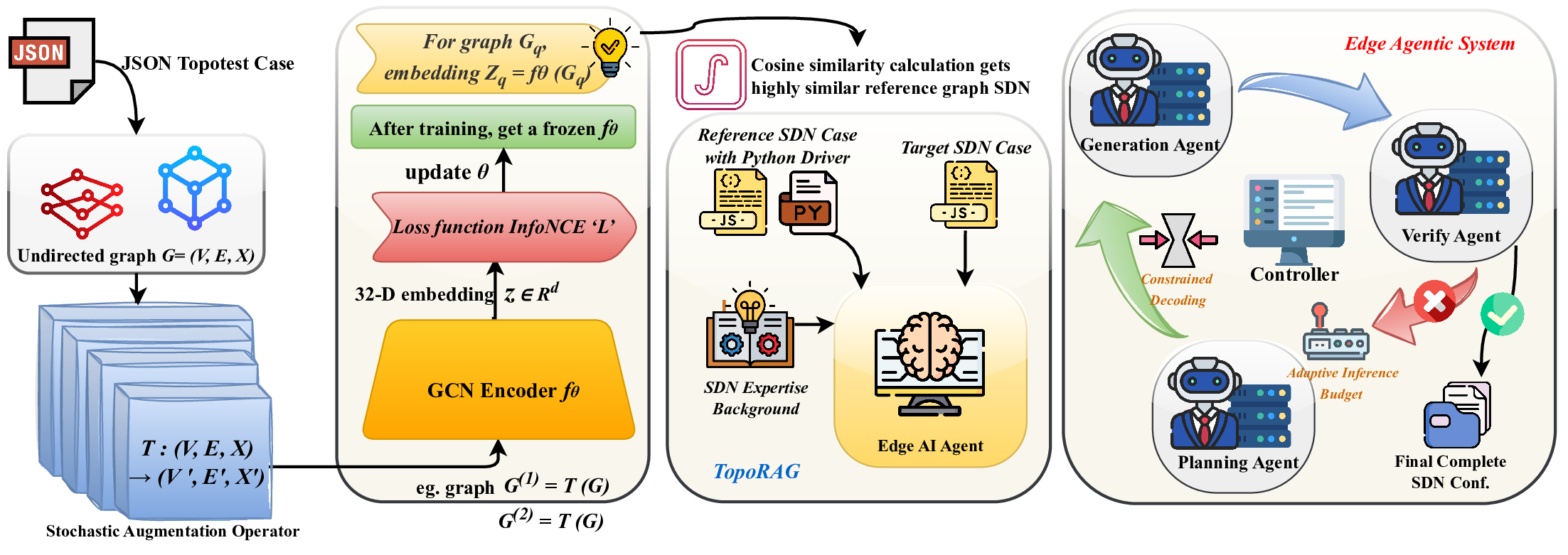}
  \caption{TopoEdge: a topology JSON is encoded and embedded for nearest-neighbour retrieval of a verified reference, forming TopoRAG that grounds a distributed AI agent planning--generation--verify loop for iterative configuration generation and repair.}
  \label{fig:framework}
\end{figure*}

From the FRRouting Topotest corpus \cite{FRRTopotests}, we extract $1{,}879$ topology JSON graphs. Among them, we identify a curated verified set $\mathcal{V}$ of $386$ cases for which each topology JSON is aligned with an executable and validated Python driver. We reserve $\mathcal{V}$ exclusively as the pool of retrieval anchors to ensure that any retrieved reference always comes with a verified driver.

To avoid data leakage between retrieval and representation learning, we remove $\mathcal{V}$ from the topology-only graph pool and split the remaining $1{,}493$ graphs into $1{,}493$ training graphs and $136$ validation graphs for self-supervised encoder training, while holding out an additional $250$ graphs as a topology-only test split.
For end-to-end evaluation, we sample a reference set $R$ of $50$ cases from $\mathcal{V}$ as retrieval candidates and select a query set $Q$ of $200$ cases from the held-out test split as evaluation targets, with $Q \cap R = \emptyset$ and all splits mutually disjoint.

TopoEdge trains a topology encoder that is label-free and structure-sensitive, so that similarity can be computed directly from topology graphs without manual annotations. A three-layer GCN is adopted as the encoder backbone because it is computationally efficient, stable on small sparse graphs, and naturally captures multi-hop connectivity via neighborhood aggregation. Three layers provide a practical depth to encode local-to-meso-scale structure (e.g., two-to-three hop neighbour) without over-smoothing on the relatively small FRRouting topologies.\footnote{GCN-style message passing is a standard and effective choice for learning on sparse graphs. Since explicit similarity labels are not available, TopoEdge uses self-supervised contrastive learning to obtain discriminative topology embeddings. This follows established InfoNCE(information noise-contrastive estimation) objectives \cite{oord2018cpc} and graph-contrastive pipelines with stochastic topology augmentations.}

For contrastive representation learning, we apply a stochastic augmentation operator \(\mathcal{T}:(V,E,X)\mapsto (V',E',X')\) that independently drops each edge with probability \(p_{\mathrm{edge}}\) and each node with probability \(p_{\mathrm{node}}\), and restricts \(X\) to the remaining nodes. Given a graph \(G\), we sample two augmented views \(G^{(1)}=\mathcal{T}(G)\) and \(G^{(2)}=\mathcal{T}(G)\). Both views are encoded by the same three-layer graph convolutional encoder \(f_{\theta}:(V,E,X)\mapsto z\in\mathbb{R}^{d}\), which outputs a fixed-dimensional graph embedding \(z\) via global pooling over node representations. We set \(d=32\) and normalise \(z\) to unit norm so that cosine similarity is well-defined and comparable across cases.

Training uses an InfoNCE contrastive objective over mini-batches. For a batch of graphs \(\{G_k\}_{k=1}^{B}\) with view embeddings \(z_k^{(1)}\) and \(z_k^{(2)}\), the loss is
\begin{equation}
  \mathcal{L}
  = - \sum_{k=1}^{B}
    \log
    \frac{
      \exp\left(\mathrm{sim}(z_k^{(1)}, z_k^{(2)})/\tau\right)
    }{
      \sum_{j=1}^{B} \exp\left(\mathrm{sim}(z_k^{(1)}, z_j^{(2)})/\tau\right)
    },
\end{equation}
where \(\mathrm{sim}(\cdot,\cdot)\) denotes cosine similarity and \(\tau>0\) is a temperature parameter. This objective pulls together embeddings of two views of the same topology while pushing apart embeddings of different topologies within the batch. After convergence on the training and validation splits, the encoder \(f_{\theta}\) is frozen and used only for retrieval.

At inference time, the frozen encoder embeds an unseen target topology and all reference topologies in \(R\). A cosine nearest-neighbour lookup selects the most similar reference case that has a verified Python driver. TopoEdge then assembles a topology-grounded retrieval-augmented context, TopoRAG, which includes: (i) the target topology JSON, (ii) the retrieved reference topology JSON, (iii) the retrieved reference verified Python driver, and (iv) SDN expertise background knowledge. This TopoRAG context becomes the shared grounding for the downstream edge-deployed agentic system in Figure~\ref{fig:framework}. The retrieved reference anchors subsequent generation and repair, while the later stages enforce correctness through a generate--verify--repair loop and regulate cost via adaptive inference budgets and constrained decoding.


\subsection{Distributed LLM Agents and Validation Loop}\label{subsec:agents_loop}

TopoEdge deploys a distributed small-model LLM on an edge cluster and exposes it as three role-specialised agents: Planning agent, Generation agent, and Verify agent. The three agents share the same quantised model runtime and weights, while their behaviours differ through role-specific system prompts, output contracts, and tool bindings. A central controller coordinates all calls. It assembles TopoRAG context, dispatches prompts to the edge cluster, materialises generated artefacts into the target case directory, and maintains iteration state across rounds.

\begin{figure}[t]
  \centering
  \includegraphics[width=0.8\linewidth]{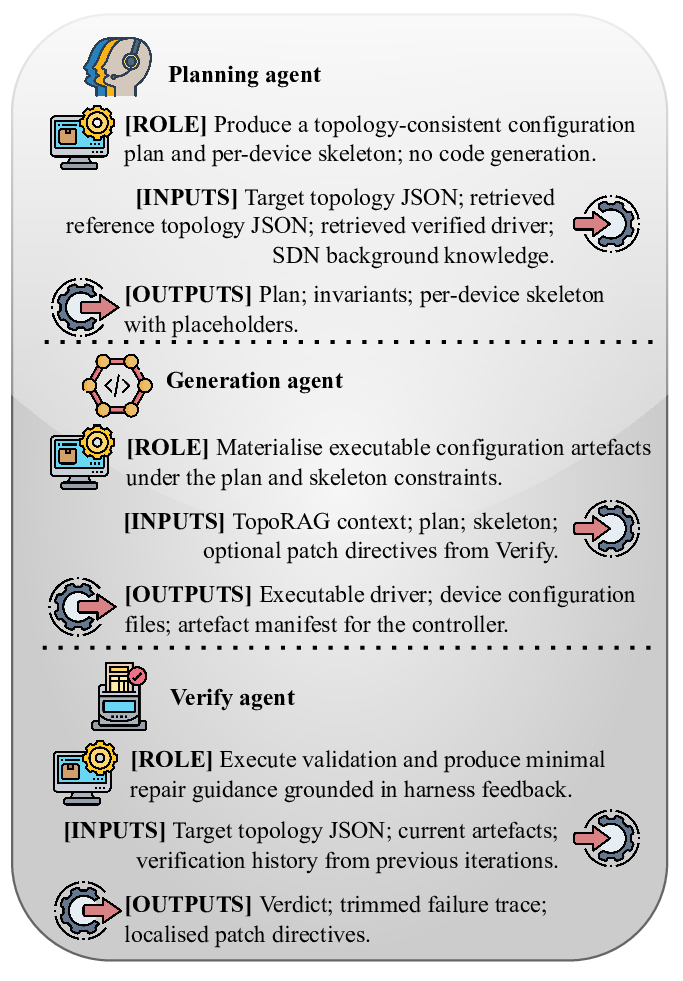}
  \caption{Example prompts for Planning, Generation, and Verify agents.}
  \label{fig:prompts}
\end{figure}

All reasoning and execution are organised as a closed loop that is grounded by TopoRAG and driven by role-separated prompting. Figure~\ref{fig:prompts} sketches the prompt layouts. The controller first invokes the Planning agent using the target topology JSON, the retrieved reference topology JSON, and the verified reference Python driver. The Planning agent returns a structured configuration plan and a per-device skeleton. The plan fixes topology-consistent protocol intent and invariants, while the skeleton defines device-level configuration blocks with placeholders. This step constrains the downstream search space and makes later repairs localised.

The controller then invokes the Generation agent with TopoRAG plus the plan and skeleton. The Generation agent produces executable configuration artefacts for the target case, including the main Python driver and, when needed, per-device configuration files. The controller writes these outputs to disk and hands control to the Verify agent. Unlike a separate validation machine, the Verify agent is responsible for running the FRRouting Topotest plus \texttt{pytest} harness, collecting verdicts, and trimming logs into a compact failure trace suitable for repair.

If verification passes, the controller finalises the current artefact and either proceeds to the next stage or terminates. If verification fails, the controller invokes the Verify agent again in repair mode, supplying the target topology JSON, the last generated artefacts, and the trimmed failure trace. The Verify agent returns minimal patch directives that identify the affected devices and configuration blocks and specify edits while preserving unchanged regions. The controller converts these directives into explicit patch instructions and resubmits them to the Generation agent for the next attempt. This generate--verify--repair cycle continues until success or an iteration budget is reached.

\subsection{Adaptive Inference Budget and Constrained Decoding}\label{subsec:budget_decode}

The three-agent loop in Section~\ref{subsec:agents_loop} is designed for edge deployment, but it still benefits from explicit controls to remain efficient and stable on a distributed small-model runtime. TopoEdge therefore adds two lightweight controllers that operate alongside the central controller and apply across the agentic loop: an adaptive inference-budget controller that limits per-case generation and iteration, and a constrained decoding layer that restricts what the Generation agent can emit when producing configuration artefacts.

The adaptive budget controller is applied at the start of each case, before invoking the Planning agent. It estimates difficulty using only signals available from TopoRAG, while targeting predictable inference cost under tight edge resources. More broadly, controlling decoding effort is a standard lever for reducing LLM inference overhead and latency \cite{leviathan2023speculative}. Let \(z_q \in \mathbb{R}^{d}\) denote the frozen GCN embedding of the target topology and \(z_r \in \mathbb{R}^{d}\) the embedding of its nearest retrieved reference. The controller computes a similarity score \(s^\star(q)=\mathrm{sim}(z_q,z_r)\), where \(\mathrm{sim}(\cdot,\cdot)\) is cosine similarity. It also extracts lightweight topology statistics from the parsed graph, including the number of nodes \(|V_q|\), the number of edges \(|E_q|\), and the maximum node degree \(\Delta_q\). These indicators are mapped to a scalar difficulty score \(d_q\in[0,1]\) via a calibrated function \(d_q=h\!\left(s^\star(q),|V_q|,|E_q|,\Delta_q\right)\), where larger values indicate harder cases. Based on \(d_q\), the controller selects a unified resource envelope for the loop, including a per-call token cap \(B_{\mathrm{tok}}\) and a maximum number of generate--verify--repair iterations \(B_{\mathrm{iter}}\). This budget is fixed per case and enforced consistently across Planning, Generation, and Verify calls, yielding predictable edge cost while allocating more computation to low-similarity or structurally complex cases.

Constrained decoding is applied when the controller invokes the Generation agent, where uncontrolled sampling is both error-prone and expensive. The constraint layer is instantiated from the Planning outputs, which provide a per-device skeleton marking fixed blocks and parameter placeholders. At each decoding step \(t\), the layer derives a permitted token set \(C_t\) from the current skeleton position. Fixed positions restrict \(C_t\) to the expected literal tokens. Placeholder positions restrict \(C_t\) using a domain schema and case-local context, such as valid interface identifiers from the topology JSON, admissible address-family formats for prefixes, and constrained lexical choices for command keywords. This form of grammar-guided restriction is closely related to grammar-constrained decoding, which is effective for enforcing structured outputs without fine-tuning. Let \(V\) denote the model vocabulary and \(p_{\mathrm{LLM}}(\cdot \mid y_{<t},x)\) the base next-token distribution given the prompt \(x\) and prefix \(y_{<t}\). The constrained distribution masks invalid tokens,
\begin{equation}
p(y_t \mid y_{<t},x)=
\frac{p_{\mathrm{LLM}}(y_t \mid y_{<t},x)\,\mathbf{1}[y_t\in C_t]}
{\sum\limits_{v\in V} p_{\mathrm{LLM}}(v \mid y_{<t},x)\,\mathbf{1}[v\in C_t]}.
\end{equation}

and decoding is performed only over \(C_t\). This concentrates model capacity on topology-dependent choices and reduces wasted tokens on illegal syntax, unsupported commands, and schema-invalid parameters.

Both controls are integrated into the same execution path as the agents rather than as post-processing. The budget controller is computed once per case from TopoRAG signals and bounds the full planning-generation-verify loop. The constrained decoding layer is activated only for Generation calls and is parameterised by Planning outputs, while the Verify agent continues to execute Topotest and request minimal patches when failures occur. Together, these components improve reliability and efficiency on the edge: easy cases converge under tight budgets, hard cases receive additional iterations, and generated artefacts remain close to valid configuration syntax and schema throughout the loop.

\section{EVALUATION AND ANALYSIS}
\label{sec:evaluation}
\subsection{Experimental Setup and Metrics}

We evaluate TopoEdge on SDN configuration generation using a held-out query set $Q$ of $200$ Topotest cases (drawn from the topology-only test split) and a reference set $R$ of $50$ verified cases (sampled from the curated verified set $\mathcal{V}$) as retrieval candidates, as defined in Section~III-A.

TopoEdge is deployed as a distributed edge agent system. LLM inference is served by an edge cluster of four Raspberry Pi nodes running Llama-3.1-8B Instruct Q40 via a local, on-device inference runtime (a llama.cpp/ GGUF (quantized model format) backend on ARM CPUs with SIMD (vector instruction set) support). The cluster acts as an inference pool: the same quantised model is replicated across nodes, and requests are scheduled over TCP for concurrency rather than model-parallel execution. The model is instantiated as three role-specialised agents---Planning agent, Generation agent, and Verify agent---that share weights but differ by system prompts, output contracts, and tool bindings. A central controller coordinates the loop, assembles TopoRAG, dispatches prompts, persists artefacts into the case directory, and tracks iteration state. The Verify agent owns execution: it runs the Topotest harness, trims logs into compact failure traces, and emits localised patch directives that are fed back to the Generation agent.

We compare TopoEdge to two controlled baselines.
\textbf{(1). No-TopoRAG} keeps the identical edge deployment, agents, and loop, but disables topology-aware retrieval so the controller provides only the target topology JSON plus fixed SDN background knowledge.
\textbf{(2). Central-LLM} keeps TopoRAG and the same agent contracts, but replaces the edge model with a centralised agent accessed via API; concretely, we use Claude Code as the inference engine while holding the tool protocol and verification procedure constant.

We report four groups of metrics. First, the cumulative pass rate versus iteration, which shows how quickly the loop accumulates solved cases. Second, the distribution of iteration-at-pass over successful cases, which characterises convergence speed conditional on success. Third, Pass@\(\{1,5,10,20\}\), which summarises success under fixed iteration budgets. Fourth, efficiency indicators: average iterations per case, average wall-clock time per case, and average generated tokens per case, all computed over the full evaluation set so that capped failures contribute to cost. Finally, we summarise failure modes among cases that exhaust the 20-iteration budget.

\begin{figure}[t]
  \centering
  \includegraphics[width=0.95\linewidth]{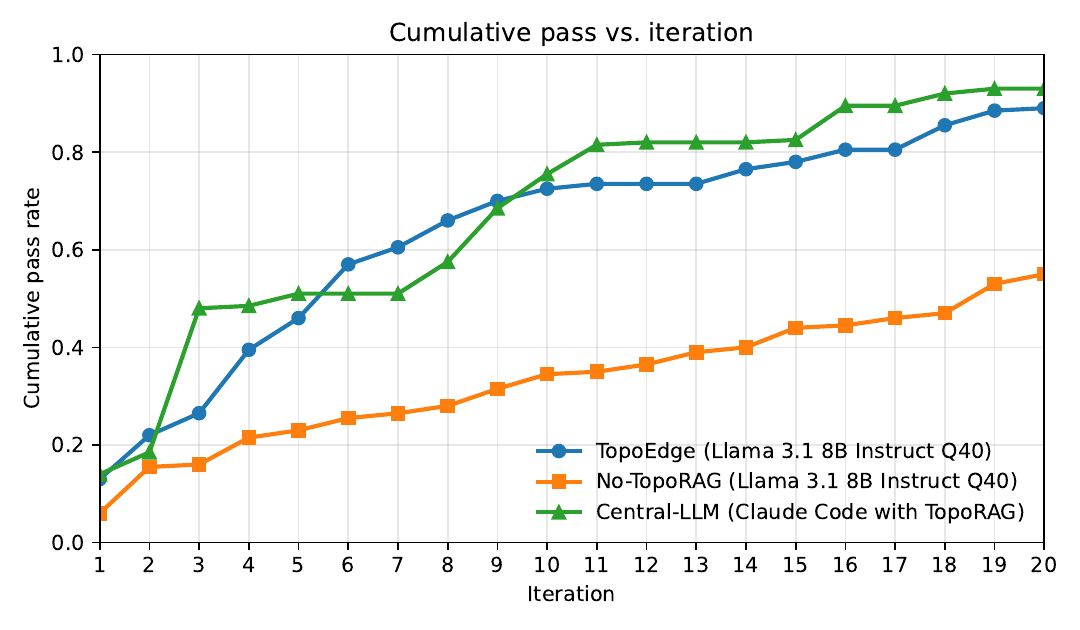}
  \caption{Cumulative pass rate versus iteration.}
  \label{fig:cum_pass}
\end{figure}

\begin{figure}[t]
  \centering
  \includegraphics[width=0.95\linewidth]{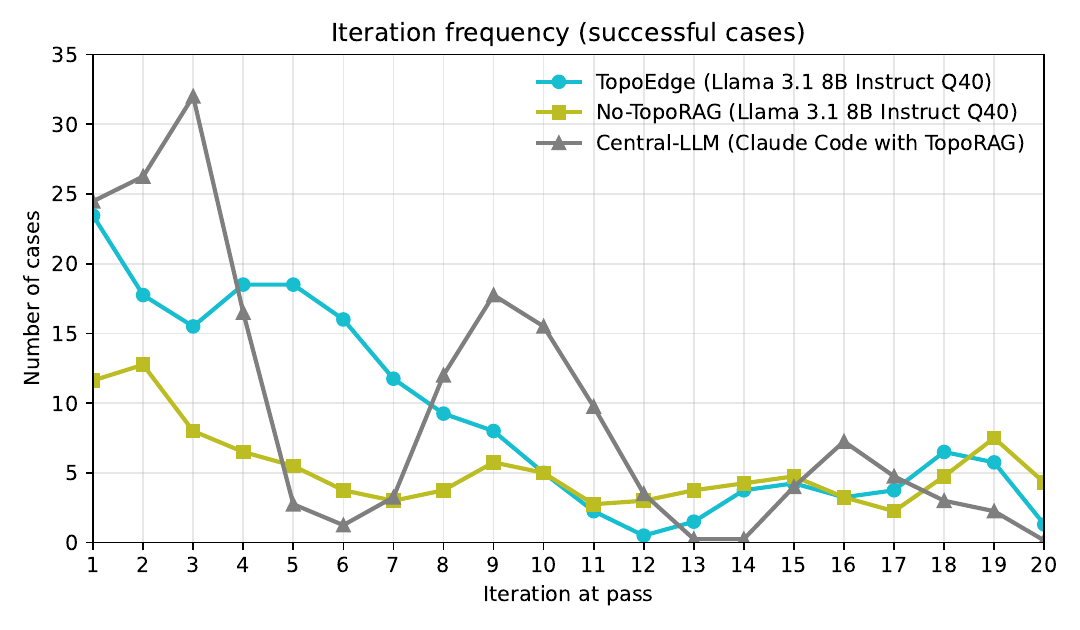}
  \caption{Iteration-at-pass frequency among successful cases.}
  \label{fig:iter_freq}
\end{figure}

\begin{table}[t]
  \centering
  \small
  \resizebox{0.92\linewidth}{!}{
  \begin{tabular}{lcccccc}
    \hline
    Method & P@1 & P@5 & P@10 & P@20 & Med & Avg \\
    \hline
    TopoEdge    & 0.130 & 0.460 & 0.725 & 0.890 & 5 & 7.835 \\
    No-TopoRAG  & 0.060 & 0.230 & 0.345 & 0.550 & 9 & 13.275 \\
    Central-LLM & 0.140 & 0.510 & 0.755 & 0.930 & 4 & 7.035 \\
    \hline
  \end{tabular}}
  \caption{Success and iteration statistics on 200 held-out cases. Med is the median iterations to pass among successful cases. Avg is the average iterations per case, including capped failures.}
  \label{tab:main_metrics}
\end{table}

\begin{table}[t]
  \centering
  \small
  \resizebox{0.92\linewidth}{!}{
  \begin{tabular}{lcccc}
    \hline
    Method & Time (s) & Tokens & Verify runs & Cap fails \\
    \hline
    TopoEdge    & 220 & 4{,}000 & 7.8  & 22 \\
    No-TopoRAG  & 360 & 6{,}900 & 13.3 & 90 \\
    Central-LLM & 75  & 4{,}400 & 7.0  & 14 \\
    \hline
  \end{tabular}}
  \caption{Efficiency metrics. Verify runs align with loop iterations. Cap fails are cases that do not pass within 20 iterations.}
  \label{tab:efficiency}
\end{table}

\begin{table}[t]
  \centering
  \small
  \resizebox{0.92\linewidth}{!}{
  \begin{tabular}{lcccc}
    \hline
    Variant & P@20 & Avg iters & Tokens & Invalid rate \\
    \hline
    Full TopoEdge                & 0.890 & 7.835  & 4{,}000 & 0.045 \\
    Without adaptive budgets     & 0.875 & 8.150  & 4{,}700 & 0.048 \\
    Without constrained decoding & 0.840 & 9.400  & 4{,}600 & 0.090 \\
    Without both                 & 0.800 & 10.200 & 5{,}400 & 0.140 \\
    \hline
  \end{tabular}}
  \caption{Ablation of system controllers. Invalid rate counts iterations that produce non-executable artefacts or schema-violating parameters.}
  \label{tab:ablation}
\end{table}
\vspace{-0.5em}

\subsection{Results and Analysis}

Figure~\ref{fig:cum_pass} reports the cumulative pass rate versus iteration. At the full 20-iteration budget, TopoEdge reaches Pass@20 of 0.890, corresponding to 178 passing cases out of 200. No-TopoRAG reaches Pass@20 of 0.550, or 110 out of 200, showing that removing retrieval substantially reduces both early success and eventual coverage. Central-LLM reaches Pass@20 of 0.930, or 186 out of 200, indicating that stronger centralised inference improves outcomes, but the remaining gap to TopoEdge is modest given that TopoEdge runs entirely on an edge cluster. Read together, these results support the core design claim: topology-grounded reference transfer is the primary driver of robustness, while the agentic loop and verification protocol stabilise synthesis under tight edge constraints.

Figure~\ref{fig:iter_freq} shows iteration-at-pass frequencies over successful cases. TopoEdge concentrates a larger fraction of its passes in earlier iterations and exhibits a shorter long tail than No-TopoRAG, consistent with TopoRAG reducing the search space by anchoring protocol structure and parameter regimes to a verified neighbour. No-TopoRAG shifts more successful cases into later rounds, reflecting repeated corrections to peering structure, addressing, and interface binding without a strong structural prior. Central-LLM shows the strongest early convergence, which aligns with its higher Pass@5 and Pass@10, but still benefits from the same TopoRAG grounding and the same execution-centric repair protocol.

Table~\ref{tab:main_metrics} quantifies the same trend as Figures~\ref{fig:cum_pass}--\ref{fig:iter_freq}: TopoRAG mainly improves reliability under limited budgets by providing a transferable structural prior. Compared with No-TopoRAG, TopoEdge achieves materially higher Pass@5 and Pass@10 and reaches success in fewer iterations, which indicates that retrieval reduces the amount of exploratory search the loop must perform before it becomes topology-consistent. Central-LLM strengthens early convergence further, but the remaining gap at the full budget is small, suggesting that once generation is grounded by a verified neighbour and corrected by execution feedback, iteration and repair structure matter nearly as much as raw model capacity.

Table~\ref{tab:efficiency} and Table~\ref{tab:ablation} clarify how TopoEdge trades off cost and stability. No-TopoRAG is expensive because verification runs are spent repairing topology-level inconsistencies that would otherwise be anchored by the retrieved reference, leading to more capped failures and longer trajectories. The ablation implies a division of labour between controllers: adaptive budgets primarily regulate cost by avoiding over-generation on easy cases, whereas constrained decoding primarily regulates correctness by preventing schema-violating artefacts that waste iterations.

\section{CONCLUSION AND OUTLOOK}
\label{sec:conclusion}
TopoEdge is a topology-driven edge framework for SDN configuration generation and repair. It integrates TopoRAG, which retrieves a verified nearest-neighbour case in a contrastively learned topology embedding space, with a distributed generate--verify--repair loop designed for resource-constrained hardware. The loop is realised by three role-specialised agents—Planning, Generation, and Verify—coordinated by a central controller. On 200 held-out Topotest cases, TopoEdge achieves Pass@20 of 0.890, substantially outperforming the No-TopoRAG variant (0.550), which confirms topology-aware retrieval as the primary source of robustness. TopoEdge also approaches a strong centralised baseline, Central-LLM (0.930), indicating that retrieval grounding and execution-centric repair can largely close the performance gap between edge-deployed small models and centralised inference. Iteration-at-pass and efficiency metrics further show that TopoRAG reduces unproductive search, while the Verify agent stabilises the loop by transforming harness feedback into compact failure traces and localised patches.

Future work will extend evaluation to larger and more heterogeneous topology distributions to stress-test generalisation and long-tail failures. We also plan to explore richer agent designs and coordination policies under the same edge constraints, including alternative role decompositions, tighter patch contracts, and improved controller-side scheduling. Finally, we will revisit the topology encoder to consider higher-capacity graph encoders and, where appropriate, topological neural network variants that can exploit higher-order motifs beyond rank-one link structure.

{\footnotesize

}

\end{document}